\documentclass[aps,prl,twocolumn,amsfonts,showpacs,superscriptaddress]{revtex4} 
\usepackage{bm}
\usepackage{epsfig,amsopn}
\usepackage{graphicx}
\usepackage{amsmath,amssymb}
\usepackage{natbib}
\renewcommand{\section}[1]{{\par\it #1.---}}
\def\beq{\begin{eqnarray}}
\def\eeq{\end{eqnarray}}

\def\bn{{\bm{n}}}
\def\be{{\bm{\hat{e}}}}
\def\bell{\bm \ell}
\def\br{\bm r}
\begin{document}

\title{Additivity Principle in High-dimensional Deterministic Systems}
\author{Keiji Saito}
\affiliation{Graduate School of Science, 
University of Tokyo, 113-0033, Japan} 

\author{Abhishek Dhar}
\affiliation{Raman Research Institute, Bangalore 560080, India}
\date{\today}

\begin{abstract}
The additivity principle (AP), conjectured by Bodineau and Derrida 
[Phys. Rev. Lett. {\bf 92}, 180601 (2004)], is discussed 
for the case of heat conduction 
in three-dimensional disordered harmonic lattices to consider 
the effects of deterministic dynamics, 
higher dimensionality, and different transport regimes, i.e., 
ballistic, diffusive, and anomalous transport. 
The cumulant generating function (CGF) for heat transfer is accurately
calculated, and  compared with the one given by the AP. 
In the diffusive regime, we find a clear agreement with the
conjecture even if the system is high-dimensional. 
Surprisingly even in the anomalous regime the CGF is also well  
fitted by the AP. Lower dimensional systems are also studied
and  the importance of three-dimensionality for the validity is stressed.
\end{abstract}

\maketitle 
Understanding the general features of 
currents and their fluctuations in transport phenomena
is one of the main goals in nonequilibrium statistical physics. 
Heat conduction is a typical transport phenomenon, where one considers 
heat transferred through a system from a bath at temperature
$T_L$ to a bath at temperature $T_R$. Transport can be classified 
by the system-size dependent thermal conductivity defined 
in the linear response regime. Considering a slab of width $W$ and length
$N$, for a small $\Delta T=T_L-T_R$ applied across its length 
and with $T=(T_L+T_R)/2$ we define 
\beq
\kappa(T) \equiv {J/W^2 \over  \Delta T/N} \propto N^{\alpha} , \label{tc}
\eeq
where $J$ is the energy current. 
Fourier's law implies $\alpha=0$, while $\alpha=1$ is ballistic transport.
Many dynamical systems with momentum conservation show 
anomalous transport ($0 < \alpha < 1$)\cite{BLR00,LLP03,dhar08}. 

In this letter, we consider
the properties of current fluctuations
beyond the linear response regime
in different transport regimes characterized by the parameter $\alpha$. 
One of the universal properties of current fluctuations which is expected to
be valid irrespective of $\alpha$, is the fluctuation theorem
\cite{evan1,gc,lebo99,Kur98}. 
It quantitatively connects distribution of positive and negative heat transfer 
and is valid in  the far from equilibrium regime. 
For diffusive systems, where Fourier's law is satisfied 
$(\alpha=0)$, some important progress has been made.
Bodineau and Derrida \cite{BD04} made a remarkable conjecture,
namely  the additivity
principle (AP), which enables one to compute 
all higher orders of current cumulants given just the temperature dependent
thermal conductivity $\kappa(T)$ of a system. 
Bertini and co-workers \cite{BSGJL02} introduced a 
macroscopic fluctuation theory (MFT) that describes the asymptotic probability 
of observing a given time-dependent local current and temperature profile. 
The MFT is 
expected to be valid for a wide class of stochastic 
models, and  the AP can be derived from it under the condition that the 
dominant trajectories are time-independent.
From this, the sufficient condition to get AP from the viewpoint of the MFT is 
given by \cite{BD04} 
\beq
\kappa(T) [ \kappa (T) T^2 ]''
 \le [\kappa (T) ]'  [ \kappa (T) T^2 ]'~. \label{sufficient}
\eeq

However, we do not still have the necessary and sufficient condition for the AP. 
One of the strategies for finding  the condition is to test the AP in different concrete models.
It was confirmed that the AP is consistent with the exact expressions of
several orders of current cumulants in the symmetric simple 
exclusion process \cite{BD04}. 
Recently, the AP was numerically verified in another stochastic system,  
namely for heat transport in the Kipnis-Marchioro-Presutti model \cite{HG09},
by measuring rare events with  a sophisticated algorithm \cite{GKP06}. 
However, studies so far are concerned only to stochastic processes
where not only the reservoir but the system dynamics also is
probabilistic.
We do not still understand the effects of
{\em deterministic system dynamics}, {\em higher dimensionality}, and 
{\em non-diffusive transport} ($\alpha \neq 0$). Hence, it is of general interest to consider the AP for systems with bulk Hamiltonian dynamics 
attached to stochastic thermal reservoirs.
In this letter, we for the first time 
address these effects using the three-dimensional mass
disordered  harmonic crystal which was recently 
shown to show different regimes of transport, 
such as ballistic ($\alpha=1$), diffusive ($\alpha=0$), and anomalous transport 
($0 < \alpha < 1$) \cite{kundu10}. 

Let $Q$ be the heat transferred from the left reservoir to the system during
the measurement time $\tau$, and  let 
$P_N(q, T_L, T_R)$ $(q\equiv Q/\tau)$ be the distribution of $Q$ for the system with
the size $N$. 
In general, the distribution at large $\tau$ has the large deviation 
form $P_N \sim e^{\tau h_N (q, T_L , T_R )}$, where $h_N$ is the large
deviation function (LDF).
The AP states that the LDF is given by the sum of the LDFs of subsystems of
length $n$ and $N-n$:  
$h_N (q, T_L , T_R) = \max_{T} \left[ h_n (q, T_L , T) + h_{N-n} (q, T,
  T_R)\right].$  
This conjecture is applied iteratively to break the system into a number of
smaller pieces. Using the assumption of local equilibrium for the small pieces
one can obtain an explicit expression for the  LDF of the full system
\cite{BD04}. 
The cumulant generating function (CGF) 
defined as 
\beq
\mu (\lambda ) &=& \lim_{\tau\to\infty}{1\over \tau}
\log \langle e^{\lambda Q} \rangle 
\eeq
is connected to the LDF through the Legendre transformation 
$\mu(\lambda ) = \max_{q} \left[ \lambda q + h_N(q, T_L ,T_R) \right]$ 
and from this one also gets an expression  for the CGF \cite{BD04}.
Verifying the AP prediction through direct simulations of heat conduction in
a $3D$ crystal is extremely difficult. In this letter we use some recent exact
results on the CGF of a harmonic crystal to test the AP.

\section{Model and Methods}
We consider a $3D$ cubic harmonic crystal with a scalar displacement 
field $x_{\bn}$ 
on each lattice
site $\bn = (n_1 , n_2 , n_3)$ where $n_1=1,2,\ldots, N$ and $n_2,\,n_3=1,\ldots W$.
The Hamiltonian is given by
\beq
{\cal H}&=& \sum_{\bn} {m_{\bn}{\dot{x}}_\bn^2 \over 2}+ {k_0 \over 2} x_\bn^2
+
 {1\over 2}\sum_{\bn,\be}
 (x_\bn-x_{\bn+\be})^2 ,
\label{ham}
\eeq
where $\be$ denotes unit vectors in the three dimensions. 
We have set the spring constant between sites to one. 
Masses are randomly distributed as
\beq
m_{\bn} = 1 - \Delta ~~~{\rm or}~~~1 + \Delta~,
\eeq
with equal probability.
Two  faces of the crystal, namely those at $n_1=1$ and $n_1=N$, have fixed
boundary conditions and   
are coupled to white noise Langevin type heat baths at temperatures $T_L$ and
$T_R \, (< T_L)$, respectively.  
In the transverse directions, periodic boundary condition is imposed.
Let ${\bm \ell}$ and ${\bm r}$ be the sites of left and right faces, namely 
${\bm \ell}=(1,n_2,n_3)$, and ${\bm r}=(N,n_2 ',n_3 ')$. 
The equations of motion of the particles are then given by
\beq
m_{\bn} \ddot{x}_\bn &=& -k_0 x_\bn - \sum_{\be} (x_\bn-x_{\bn+\be}) \nonumber \\
&-&\sum_{\bell} \delta_{\bn, {\bell}} (\gamma {\dot{x}}_{\bell} - \eta_{\bell} )
-\sum_{\br} \delta_{\bn, {\br}} (\gamma {\dot{x}}_{\br} - \eta_{\bm r} ). ~~~\label{eom}
\eeq
The noise terms at different sites are uncorrelated, while at a given site the
noise strength is specified by the correlations $
\langle \eta_{\bell} (t) \eta_{\bell } (t' ) \rangle 
= 2 \gamma  \, T_L \delta(t -t'  ) $
and $\langle \eta_{\br} (t) \eta_{{\br}} (t' ) \rangle 
= 2 \gamma  \, T_R  \delta(t -t'  )$ 
where we have set the Boltzmann
constant to the value one. 

We assume that the initial state at $t=0$ is chosen from the steady state
distribution~. 
The heat $Q$ flowing from the left reservoir into the system
between the times  $t=0$ to $t=\tau$~ is given by 
$Q=\sum_{\bell} \int_0^\tau dt~ \dot{x}_{\bell}~(-\gamma \dot{x}_{\bell} +
\eta_{\bell})$.  
The average current $J=\langle Q \rangle / \tau $ in the harmonic crystal is given by a
landauer-like  formula \cite{segal} and this gives the following expression for
the size-dependent thermal conductivity defined in Eq.~(\ref{tc}):
\beq
\kappa &\equiv & {J/W^2 \over \Delta T/N} = {N \over 2\pi W^2} \int_{0}^{\infty} 
d\omega \, {\rm Tr} [{\cal T} (\omega ) ] , \label{landauer} \\
\left[ {\cal T} (\omega )\right]_{ {\bm \ell} , {\bm \ell}'} \! &=&\! 
4[{\bm G}^+ {\bm \Gamma}_R {\bm G}^- {\bm \Gamma}_L ]_{{\bm \ell}, {\bm \ell}'} ~, 
 \nonumber \\
{\bm G}^{\pm} (\omega ) & =& \left[ -{\bm M} \omega^2 +{\bm K} \mp 
  {\bm \Sigma}_L (\omega ) \mp {\bm \Sigma}_R (\omega ) \right]^{-1} , \nonumber
\eeq
where ${\cal T}(\omega)$ is the 
transmission matrix which 
describes transmission of phonons emitted
from a site on one face attached to 
a reservoir to a site on another face, and is a $W^2\times W^2$ matrix.
The Green's function ${\bm G}^{\pm} (\omega )$ 
is a $NW^2 \times NW^2$ matrix given by the mass and force constant matrices ${\bm M}$ and ${\bm K}$, and the self-energy matrices ${\bm \Sigma}_{L/R} (\omega )$
 whose matrix elements are
$[{\bm \Sigma }_{L/R}]_{\bn , \bn '} (\omega )= 
i \gamma \omega \delta_{\bn , \bn '}\delta_{\bn,\bell /\br }\,$. 
The matrix ${\bm \Gamma}_{L/R}
= {\rm Im}\{{\bm \Sigma}_{L/R} (\omega )\}$.

\begin{figure}
\includegraphics[width=7.5cm]{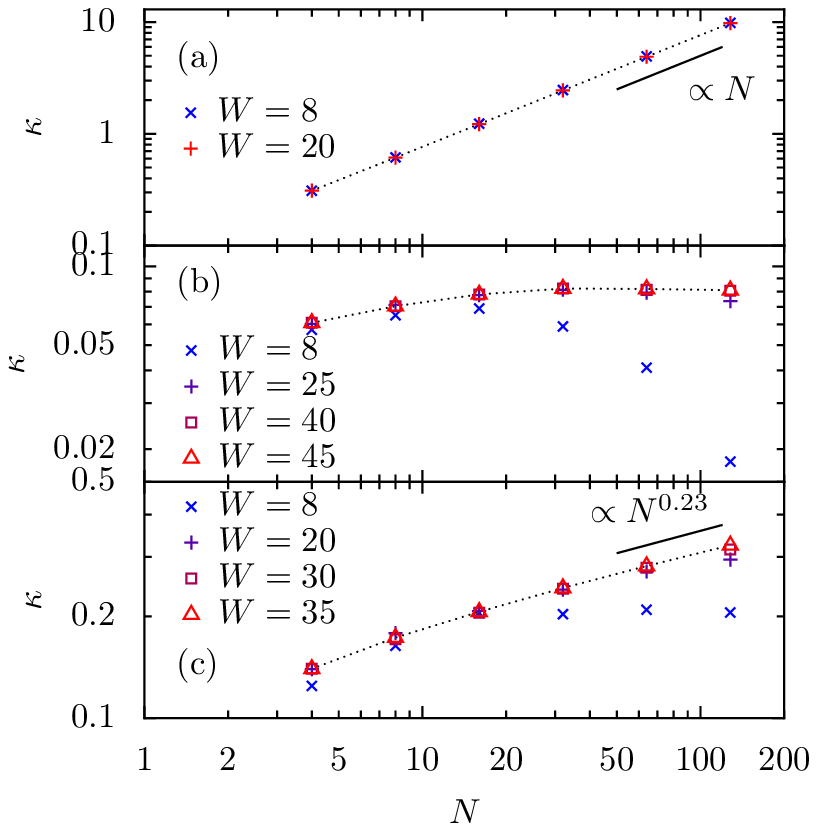}
\caption{The size-dependent thermal conductivity $\kappa$ for
$\,(\Delta, k_0)=$ (a):$\,(0,0)$, (b):$\,(0.22,9.0)$, and (c):$\,(0.82,0)$. 
For sufficiently  large $W$ we see respectively (a): ballistic ($\alpha=1$), (b): diffusive
($\alpha=0$), and (c): anomalous transport($\alpha\sim 0.23$) ~. 
Each data-point is obtained from simulations of Eq.~(\ref{eom}),
while the black dotted line is given by Eq.~(\ref{landauer}) for largest $W$.} 
\label{fig1}
\end{figure}  

Recently it was shown \cite{SD11} that, not just the current, but an exact
expression for the full CGF of the harmonic crystal can be 
obtained in terms of ${\mathcal{T}}(\omega)$ and is given by: 
\beq
\mu_{HC}(\lambda ) = \!\!
\int_{0}^{\infty} {d\omega \over 2\pi} \, {\rm Tr}
\log \Bigl[ {\bm 1}
- {\cal T} (\omega ) T_L T_R 
\lambda \left( \lambda + \Delta \beta \right)  \Bigr]^{-1}\!\!\!\!\!\! ,~\, \label{cgf}
\eeq
where $\beta_L=T_L^{-1},\beta_R=T_R^{-1}$ and $\Delta \beta=\beta_R-\beta_L$.
One can verify that the current in Eq.~(\ref{landauer}) is 
given by $J={\partial \mu_{ HC} \over \partial \lambda }\, \vline_{\lambda=0}$.  
Also Eq.~(\ref{cgf}) satisfies the fluctuation
theorem symmetry relation: $\mu (\lambda ) = \mu ( -\lambda -\Delta\beta )$ \cite{gc,SD07}.

We next discuss the prediction for CGF from the AP, which we will denote by
$\mu_{AP}(\lambda)$. 
In general the CGF can be  expressed completely as a temperature integral over
the range $[T_R,T_L]$ involving the single  parameter $\kappa(T)$. 
For the harmonic case, $\kappa$ is independent of temperature, 
and explicit expressions can be obtained for $\mu_{AP}(\lambda)$ \cite{HG09}
in terms of a single parameter $\kappa$.  
These expressions are somewhat lengthy to state and hence we give them 
in the supplementary material \cite{suppl}. 
We note that $\mu_{AP}(\lambda)$ also satisfies the symmetry: $\mu (\lambda ) = \mu ( -\lambda -\Delta\beta )$.

The main aim of this letter is to compare the AP prediction for
$\mu_{AP}(\lambda)$ with the  numerical result for $\mu_{HC}(\lambda)$
from Eq.~(\ref{cgf}).  
We note that for a $3D$ disordered crystal 
the heat current depends on the particular realization of
disorder, however for large $N$ and $W$ there is self-averaging and sample-to-sample
fluctuations become very small. Hence for a fixed disorder strength we get a
unique current and 
$\kappa$ from Eq.~(\ref{landauer}).
This value of $\kappa$ is then used to get $\mu_{AP}(\lambda)$ and compared with
$\mu_{HC}(\lambda)$.  

\begin{figure}
\includegraphics[width=7.5cm]{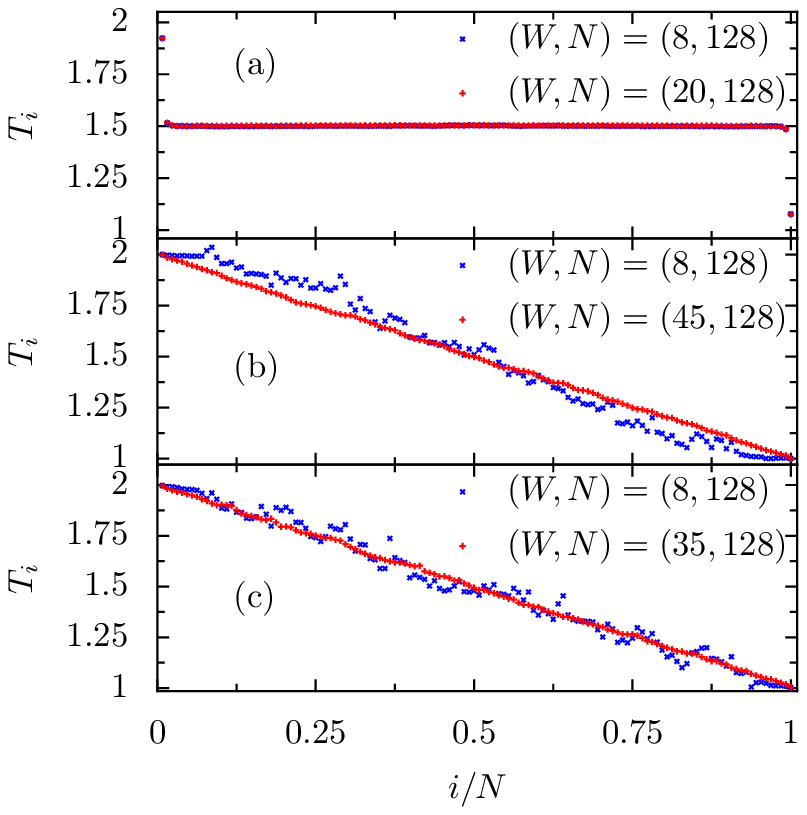}
\caption{Temperature profiles for each case in Fig.~\ref{fig1}.
Temperature is defined as $T_i =\sum_{\bn : n_1 =i} \langle m_{\bn} 
\dot{x}_{\bn}^2  \rangle / W^2$.}
\label{fig2}
\end{figure}  
\section{Average heat conduction for different regimes}
In $3D$ disordered systems without pinning potentials ($k_0=0$), 
low-frequency extended modes with diverging phonon
mean-free-paths exist and lead to anomalous transport. However,  
a  pinning potential removes these modes and transport is then governed by the 
high-frequency extended diffusive modes. Hence a $3D$ disordered pinned
crystal shows diffusive heat conduction. 
Based on the results of  \cite{kundu10} we expect different regimes of
transport and accordingly we chose the following three parameter sets for
these regimes: 
(a) ordered unpinned lattice ($\Delta =0, k_0=0$) for ballistic
transport ($\alpha=1$), (b) disordered pinned lattice ($\Delta = 0.22,
k_0=9.0$) for diffusive transport ($\alpha=0$), and  
(c) disordered unpinned lattice ($\Delta= 0.82,k_0= 0$) for anomalous
transport ($0<\alpha <1$).  
To demonstrate the different transport regimes we show the size-dependence of 
thermal conductivity in Fig.~\ref{fig1}, and  the typical temperature profiles
in Fig.~\ref{fig2}.  
The average heat current was obtained either by direct nonequilibrium
simulations of the Langevin equations Eq.~(\ref{eom}),
or from Eq.(\ref{landauer}) using 
recursive Green's function techniques \cite{kundu10} to evaluate the
transmission matrix ${\cal T}(\omega )$.
Agreement between the two methods is excellent. 
Each point in Fig.~\ref{fig1} is for one disorder realization and we fixed
parameters  $\gamma=1$, $T_L=2.0$, and $T_R=1.0$.

\begin{figure}
\includegraphics[width=7.5cm]{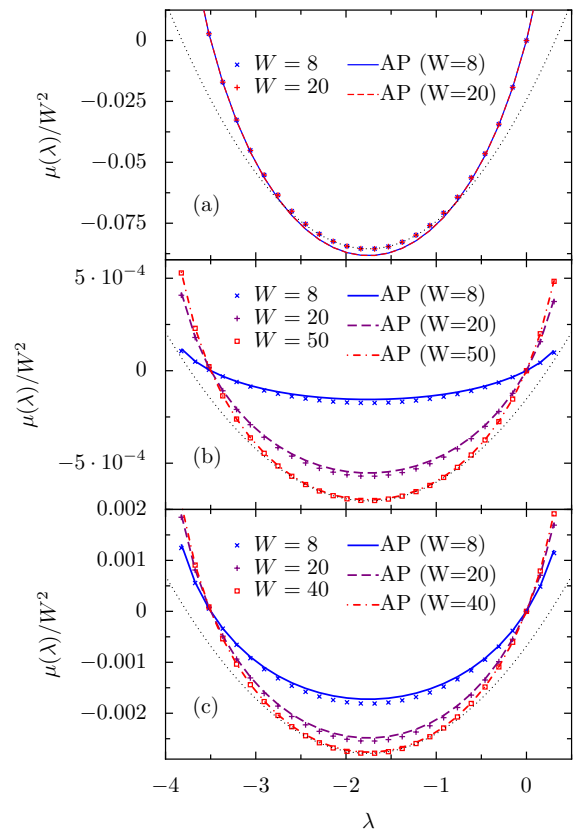}
\caption{Comparison of the numerically evaluated CGF $\mu_{HC} (\lambda )$  (points) for
  a $3D$ 
  harmonic   crystal and the AP predicted curve $\mu_{AP}(\lambda)$  for $N=128$
  and different widths. System parameters are the same as in Fig.~\ref{fig1} with 
$T_L=2.0$ and $T_R=0.25$ . The black dotted line is a quadratic fit to 
$\mu_{HC} (\lambda )$ for the largest $W$. The range of $\lambda$ is $(-\beta_R,\beta_L)$.}
\label{fig3}
\end{figure}  
Remarks on Figs.~\ref{fig1} and \ref{fig2} are in order.
From Fig.~\ref{fig1}(a)  we see that $\kappa$ is independent of width and
diverges linearly with $N$ implying ballistic transport. 
On the other hand  in Fig.~\ref{fig1}(b) and \ref{fig1}(c), we see that for small $W$,  
$\kappa$ decreases for increasing $N$. This implies the emergence of phonon
localization since the system is quasi-one dimensional.
For increasing $W$ with fixed $N$, the data converges to a constant value,
which  implies the self-averaging effect in disordered systems.
Hence, one can get precise $3D$ behavior for sufficiently large $W$. 
Fig.~\ref{fig1}(b) shows diffusive transport ($\alpha=0$) for sufficiently large $W$
and $N$, while Fig.~\ref{fig1}(c) shows anomalous 
behavior with systematic power law divergence  ($\alpha \approx 0.23$). 
The temperature profile  in Fig.~\ref{fig2}(b) shows clear linear profile 
consistent with the Fourier's law. 
Interestingly, even the anomalous  case in Fig.~\ref{fig2}(c) shows a 
linear profile which is very  different  from  
other nonlinear systems with anomalous transport, such as the Fermi-Pasta-Ulam (FPU) chain 
which have  nonlinear temperature profiles even for small temperature
difference between reservoirs \cite{LLP03,dhar08}.

\section{The CGF}
We now present results  comparing $\mu_{AP}(\lambda)$ with
$\mu_{HC}(\lambda)$ in the three different regimes~.
We obtained $\mu_{AP}(\lambda)$  by using $\kappa$ from Eq.~(\ref{landauer})
while  
$\mu_{HC}(\lambda)$ was computed from Eq.~(\ref{cgf}). 
In these computations, we set the parameters $\gamma=1.0,
T_L=2.0,T_R=0.25$  and size $N=128$. 
The results for ballistic, 
diffusive, and anomalous cases are shown in  Figs.~\ref{fig3}(a), \ref{fig3}(b),
and \ref{fig3}(c) respectively.
Note that both $\mu_{HC} (\lambda)$ in Eq.(\ref{cgf}) and $\mu_{AP}(\lambda )$ in 
\cite{suppl} 
exactly satisfy $\mu (0) = \mu (\Delta\beta)=0$.
For ballistic case (a), we see 
deviations from AP curve irrespective of $W$ as expected.
On the other hand, in Fig.~\ref{fig3}(b), we see that the agreement
between $\mu_{AP}(\lambda)$ and $\mu_{HC}(\lambda)$ improves 
for increasing $W$. For small $W$ where localization effect is dominant, 
there are clear deviation from AP curve. 
For other cases with larger temperature differences, we obtained 
agreement with the AP for sufficiently large system size \cite{suppl}.
We should also note that the present situation does not satisfy 
the sufficient condition from the MFT (\ref{sufficient}), since the thermal conductivity 
is independent of temperature. Hence, this model extends the sufficiency 
condition for the AP.
Surprisingly, for the case of 
anomalous transport, we see from Fig.~\ref{fig3}(c) behaviour
similar to the diffusive case, with clear verification of AP at sufficiently 
large $W$. Note that the original theory of the AP conjecture \cite{BD04} assumes
diffusive transport. 

\begin{figure}
\includegraphics[width=7.5cm]{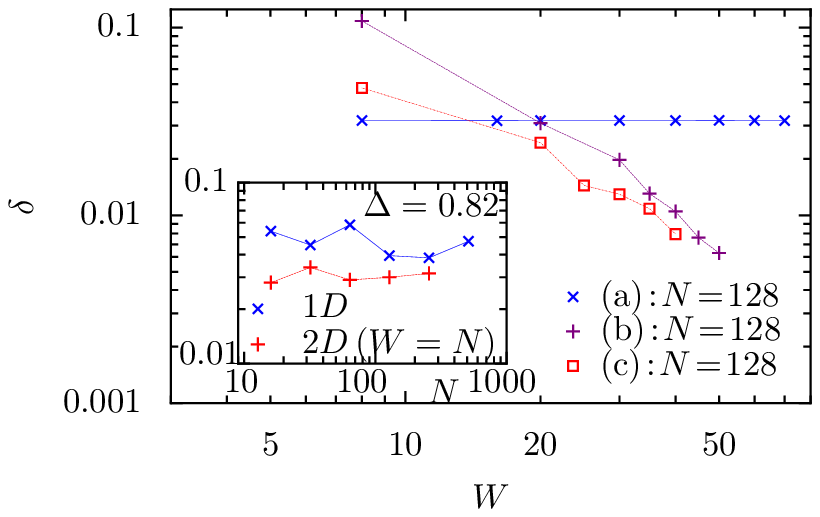}
\caption{The quantitative estimation of the degree of agreement of the CGF
  with AP. The inset shows the results for low-dimensional anomalous cases.
$(T_L,T_R)=(2.0,0.25).$} 
\label{fig4}
\end{figure}

The degree of coincidence seen in Fig.~\ref{fig3} is now quantitatively
discussed. 
We note that  harmonic lattices, 
in not only $3D$, but also in $1D$ and $2D$ can show diverging thermal
conductivity. 
For instance, $1D$ disordered harmonic chains with open boundary
condition show diverging  
thermal conductivity with the power $\alpha=1/2$ \cite{MI70}.
Then, one may ask if low dimensional anomalous transport satisfies AP or not. 
Hence, in addition to $3D$ cases in Fig.~\ref{fig3}, we also discuss
low dimensional harmonic systems.  
We define the following quantity
\beq
\delta &\equiv& \Bigl|  
{\mu_{HC} (\lambda^{\ast}) - \mu_{AP} (\lambda^{\ast}) \over
\mu_{HC} (\lambda^{\ast })}  \Bigr| \, , 
~~~~~~~~~\lambda^{\ast}=-{\Delta\beta \over 2}\,,~~~~
\eeq 
where $\lambda^{\ast}$
is the value of $\lambda$ which  minimizes $\mu_{AP} (\lambda)$. As seen in  Fig.~\ref{fig3}, the 
deviation becomes maximum at the minimum value of the CGF.
Hence, the function $\delta$ quantitatively estimates the degree of discrepancy.
In Fig.~\ref{fig4}, we show $\delta$ as a function of $W$ for the three cases in $3D$.
Systematic approach to the AP is seen on increasing $W$ for both diffusive and
anomalous cases.  
In the inset, $1D$ and $2D$ results for $\delta$ are shown.
We consider the system size $N$ for $1D$ 
and $N\times N$ for $2D$ with open boundary condition, and hence, the $x$-axis
is $N$ (not $W$). 
For both $1D$ and $2D$, results for one realization of random mass are shown. 
Contrary to what happens in $3D$, in low dimensions we see  no sign of
decay of $\delta$, and  it remains almost constant value. This implies that
the coincidence for anomalous transport  is true only for $3D$ systems.

\section{Discussion}
We have discussed the additivity principle in high-dimensional
deterministic systems and  considered 
the effects of deterministic dynamics, dimensionality, and 
non-diffusive transport ($\alpha \ne 0$). 
The AP was originally proposed for $1D$ diffusive systems. 
Our main result  is to show the validity of  AP in a $3D$
Hamiltonian system in both the diffusive and anomalous 
regimes. 

In disordered harmonic crystals in $1D$ and $2D$ we find that AP is not
satisfied for the anomalous cases where $ 0< \alpha <1$. 
The major difference between $3D$ and 
lower dimensions
is that in $3D$ only a small fraction of the normal modes are localized while in
lower dimensions, most of the modes are localized \cite{kundu10,nagel84}.
Hence it is expected that there is no local equilibration in
low-dimensions.  
Then neither the MFT nor the AP are satisfied in these regimes.
Our study thus suggests that only the assumptions of (i) local
equilibration and (ii) a small current (requiring $\alpha <1$) are 
necessary for the validity of AP.  
The realization of  AP in diffusive and anomalous cases should be 
related to the time-independence in dominant trajectories in the MFT \cite{BSGJL02}. 
A verification for this would be an important future problem, but it is 
not possible within the present approach, and would require simulations 
with algorithms as in \cite{GKP06}.

The mechanism of 
anomalous transport in disordered harmonic lattices 
has some crucial differences from that in low-dimensional nonlinear systems 
such as FPU chain where Levy type of energy diffusion is seen \cite{zaburdaev} 
and temperature profile is always nonlinear. For disordered harmonic crystals 
the linear temperature profiles suggest that a local response relation 
$j(x) =-\kappa dT/dx$ is always valid, but with a size-dependent $\kappa$ in 
the anomalous case. On the other hand  levy type of diffusion seen in one-dimensional nonlinear  models means that the response is non-local (in space) and 
the AP may need modification.  
Hence, it will be of great interest to understand the general features of CGF
in low-dimensional nonlinear system.

KS was supported by MEXT (23740289).
AD thanks DST for support through the Swarnajayanti fellowship.

\newpage
\begin{widetext}
\begin{center}
{\bf \large Supplementary Material for \\
Additivity Principle in High-dimensional Deterministic Systems}
\end{center}
\end{widetext}

\section{Cumulant generating function from the additivity principle}
We derive the explicit
expression of cumulant generating function (CGF) 
from the additivity principle (AP) for the 
regime $-\beta_R \le \lambda \le \beta_L$. 
The AP assumes that 
the large deviation function (LDF) for the system with the size $N$ is given by \cite{sBD07}
\beq
h_N (q , T_L , T_R ) &=& -  \min_{T(x) } {1\over N} \int_{0}^{1} d x
{  \left[ N q + \tilde{\kappa} (T) {dT(x) \over d x}  \right]^2 
\over 4 T^2 \tilde{\kappa} (T)  } , \nonumber \label{ap1} \\
\eeq
where $\tilde{\kappa} (T) = W^2 \kappa (T)$. Variational problem in Eq.(\ref{ap1}) is reduced to finding the optimal profile $T(x)$
satisfying \cite{sBD07}
\beq
\left( {dT(x) \over dx } \right)^2 &=&
{(Nq)^2 \left[ 1 + 4 K T^2 \tilde{\kappa}(T) \right] \over \tilde{\kappa}^2 (T) } ,
\eeq
where the function $K$ is determined from the boundary condition $T(0)=T_L$ and $T(1)=T_R$. From now on, we consider the case of temperature-independent thermal conductivity
$\tilde{\kappa} (T) = \tilde{\kappa}$, which is the case for harmonic crystals.
We then derive the explicit expression 
of the CGF. The CGF is given by the Legendre transformation:
\beq
\mu (\lambda) &=& \max_{q} \left[ q \lambda + h_N (q, T_L , T_R )\right] .
\eeq

Suppose $T_L > T_R$ and the deviations are not too large so that the 
optimal profile remains monotonic. Then  
\beq
{dT(x) \over dx } &=&
{-Nq \sqrt{ 1 + 4 K T^2 \tilde{\kappa } } \over \tilde{\kappa} } .
\eeq
In this case, from the Legendre transformation, the CGF is given by \cite{sBD07}
\beq
\mu_{ AP} (\lambda ) &=& - {K\over N}  
\Bigl[
 \int_{T_R}^{T_L} dT 
{ \tilde{\kappa} \over 
 \sqrt{1 + 4 K T^2 \tilde{\kappa }} } 
\Bigr]^2  \, , \label{ap21}\\
\lambda &=& 
\int_{T_R}^{T_L} dT {1\over 2 T^2} 
\Bigl[ 
{ 1 \over 
 \sqrt{1 + 4 K T^2 \tilde{\kappa }} }  -1
\Bigr] \, . \label{ap22}
\eeq
This expression is valid for $\lambda_- \le \lambda \le \lambda_+$ where
$\lambda_{\pm}=(\beta_L - \beta_R \pm \sqrt{\beta_R^2 - \beta_L^2 })/2$.
To go beyond this regime of $\lambda$, we consider a nonmonotonic optimal profile 
given by 
\beq
{dT(x) \over d x}
&=& 
\left\{
\begin{array}{ll}
\pm N q { \sqrt{1 + 4 K T^2 \tilde{\kappa} }  \over \tilde{\kappa} } & ~~~~~~~0 \le x \le x_c \, , \\
\mp N q { \sqrt{1 + 4 K T^2 \tilde{\kappa} }  \over \tilde{\kappa} } & ~~~~~~~x_c  \le x \le 1 \, ,\\
\end{array}
\right.
\eeq
where $x_c$ satisfies $dT(x_c)/dx=0$ implying $1+4 K \tilde{\kappa} T^2 (x_c) =0$. In these cases,
the CGF is given by 
\beq
\mu_{ AP} (\lambda ) &=& {- K \over N}
\Bigl[ \int_{T_L}^{T(x_c) } dT {\tilde{\kappa} \over \sqrt{  1 + 4 K T^2 \tilde{\kappa} }}
  \nonumber \\
&+&  \int_{T_R}^{T(x_c) } dT {\tilde{\kappa} \over \sqrt{  1 + 4 K T^2 \tilde{\kappa} }} 
\Bigr]^2 \, , \label{ap31}\\
\lambda &=& 
 \int_{T_L}^{T(x_c) } dT {1\over 2 T^2} 
\Bigl[ 1 \pm 
{ 1 \over 
 \sqrt{1 + 4 K T^2 \tilde{\kappa }} }  
\Bigr] \nonumber \\
&+& 
 \int_{T_R}^{T(x_c) } dT {1\over 2 T^2} 
\Bigl[ -1 \pm
{ 1 \over 
 \sqrt{1 + 4 K T^2 \tilde{\kappa }} } 
\Bigr] \, . \label{ap32}
\eeq 
These expressions are valid for the regime $\lambda_+ \le \lambda \le \beta_L$ and
$-\beta_R \le \lambda \le \lambda_-$, respectively.

Simplifying Eqs. (\ref{ap21}), (\ref{ap22}), (\ref{ap31}) and (\ref{ap32}) is  
straightforward.  By evaluating the integrations we get the following explicit
expressions: 
\beq
\mu_{ AP}(\lambda ) =
\left\{
\begin{array}{l}
 -{\tilde{\kappa}\over 4N} 
\Bigl[ 
\log 
\left( {\sqrt{1 + 4 \tilde{\kappa} K T_L^2} + \sqrt{4\tilde{\kappa} K T_L^2 } 
\over  \sqrt{1 + 4 \tilde{\kappa} K T_R^2} + \sqrt{4\tilde{\kappa} K T_R^2 } } \right) 
\Bigr]^2 , \\
~~~~~~~~~~~~~~~~~~~~~~~~~~~~\cdots
\lambda_- \le  \lambda \le \lambda_+ , \\
\\ 
{\tilde{\kappa} \over 4 N} \Bigl[ \pi - (\theta_L + \theta_R ) \Bigr]^2 
~~ \cdots~  {-\beta_R \le \lambda \le \lambda_- \atop
\lambda_+ \le \lambda \le \beta_L } \, ,
\end{array} \right.  \label{apap}
\eeq
where $\theta_{\alpha}$ $(\alpha=L,R)$ is given by
\beq
\cos\theta_{\alpha}&=& \sqrt{1+ 4 \tilde{\kappa}\, K \, T_{\alpha}^2} \, , \\
\sin\theta_{\alpha}&=& \sqrt{4 \tilde{\kappa}\, |K| \, T_{\alpha}^2} \, .
\eeq
The function $K$ is given by  
\beq
K(\lambda , T_L , T_R) &=& {1 \over 16 \tilde{\kappa} } \Bigl[ 
\left( \beta_L - \beta_R -2 \lambda \right)^2 
-2\left( \beta_L^2 + \beta_R^{2 } \right) \nonumber \\
&&
+ \left( { \beta_L^{2} - \beta_R^{2} \over  \beta_L - \beta_R -2\lambda } \right)^2 
\Bigr] .
\eeq

\section{Effects of small system sizes and large temperature differences}
\begin{figure}
\includegraphics[width=8.3cm]{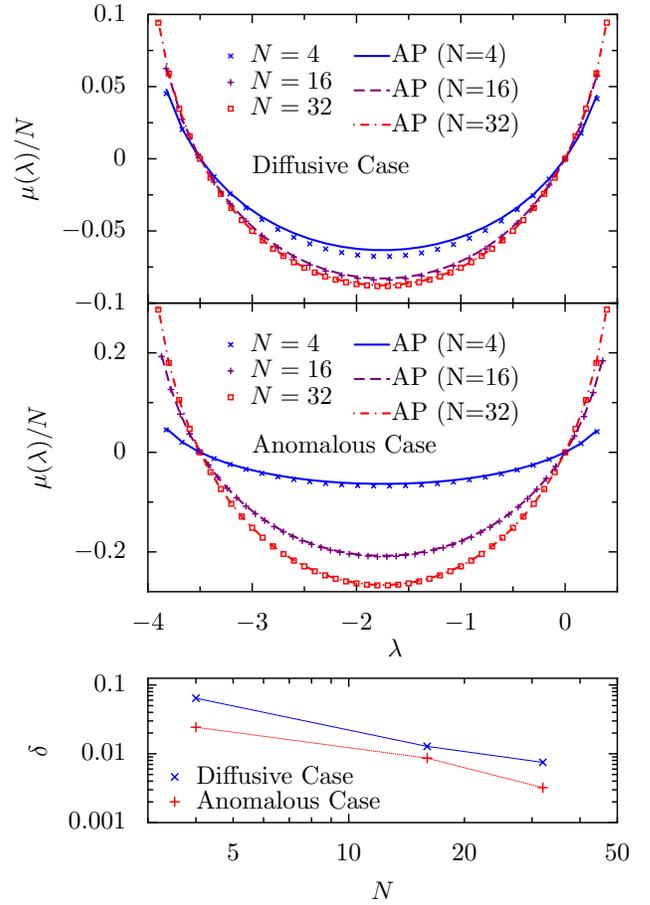} \\
\caption{The CGF in small systems. The system is a cubic structure $N^3$.}
\label{supp1}
\end{figure}

Since the AP is derived under the assumption of local equilibrium as in Eq.(\ref{ap1}),
one expects that smaller systems show larger deviation from the AP due to violation of local 
equilibrium. To see this, we consider $3D$ systems with cubic $N^3$ structure, 
and calculate the CGF for
$N=4,16,32$ with $(T_L,T_R)=(2.0,0.25)$. In Fig.\ref{supp1}, 
the CGF and the deviation $\delta$ defined 
in the main text are shown for these cases. As expected, $N=4$ clearly shows deviation from the 
AP curve. As increasing $N$, the deviation $\delta$ decreases. 

\begin{figure}
\includegraphics[width=8.0cm]{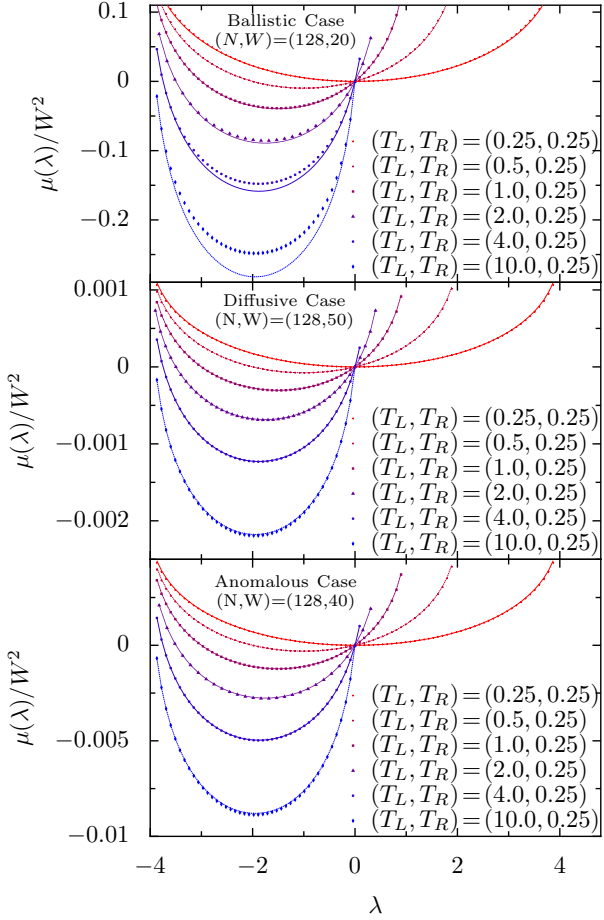} \\
\caption{The CGF in various temperature sets.
Points are $\mu_{HC}/W^2$, while solid lines are $\mu_{AP}/W^2$. 
The regime of $\lambda$ is $\lambda \in [-\beta_R , \beta_L]$}
\label{supp2}
\end{figure}

\begin{figure}
\includegraphics[width=8.0cm]{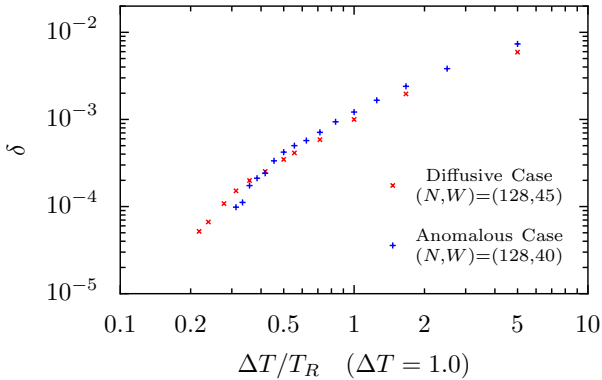} \\
\caption{Deviations from the AP in various temperature regime 
with fixed temperature difference $\Delta T =T_L-T_R=1.0$.}
\label{supp3}
\end{figure}

We next consider the effect of large temperature differences for large systems. 
For this we obtain the CGF for many temperature sets 
at fixed large $N$. We note the following exact scaling relation that holds for the CGF :
\beq
\mu (\lambda , \nu T_L , \nu T_R ) &=& \mu ( \nu \lambda , T_L , T_R ) ,
\eeq 
where $\nu$ is an arbitrary real number. This holds 
for both $\mu_{AP}$ and $\mu_{HC}$. From this relation, it is clear that if 
agreement between the formulas is seen for some temperature difference 
$ \Delta T= T_L - T_R \,(T_L > T_R)$ and average temperature $T=(T_L+T_R)/2$, then it also holds for  $\nu \Delta T$ and $\nu T$. 
Hence the correct relevant parameter is the ratio $\Delta T/T$. 
In Fig.\ref{supp2}, we show the CGF of many temperature sets with $N=128$. 
Points are $\mu_{HC}/W^2$, while solid lines are $\mu_{AP}/W^2$, and the regime of $\lambda$ is 
$\lambda \in [-\beta_R , \beta_L]$.
We see that for the ordered lattice the disagreement with AP becomes large 
on increasing $\Delta T/ T $ while for the disordered case AP is always 
satisfied.  
Fig.\ref{supp3} clearly shows that in sufficiently large systems,
the AP is accurate over a large range of $\Delta T/T$ and the agreement 
is better for smaller $\Delta T/T$. The relevance of the parameter 
$\Delta T/T$ can  be roughly understood by considering the criterion for local 
thermal equilibrium. If the typical mean free path of the heat carriers (phonons here) is denoted by $\ell_p$ then the condition for  local equilibrium is 
$\ell_p |dT/dx|/ T << 1$ \cite{kreuzer} or $\Delta T/T << L /\ell_p$. For the 
ordered  ballsitic  case $\ell_p \sim L$ and so we require $\Delta T /T << 1$. 
On the other hand for disordered systems  $\ell_p$ is finite and hence for 
sufficienlty large size $L$ the condition for local equilibrium is always
satisfied for any given $\Delta T/T$. We also note that in diffusive systems 
the condition for local equilibrium also ensures that temperature profiles do 
not show any jumps at the boundaries.

\section{Cumulant generating function for ordered harmonic crystal}
We here give the simplified expression of $\mu_{ HC} (\lambda )$ for ordered
harmonic crystal ($\Delta=0$). 
The simplified expression for $\mu_{ HC} (\lambda )$ is much more
computationally efficient than using the recursive  
Green's function technique for solving the CGF. 

Each lattice is labelled by the vector $\bn = n_1 \be_1 + n_2 \be_2 + n_3
\be_3$ where $n_1=1,...,N$ and $n_2,n_3=1,...,W$. Let $U^{( \alpha )}$ be a
matrix which acts only on the components in the $\be_{\alpha}$-direction. 
We introduce the orthogonal matrix 
\beq
U &=& {\bm 1}^{(1)}\otimes U^{(2)} \otimes U^{(3)} , \\
U^{(\alpha)}_{m,n} &=& \sqrt{2\over W} \cos \left( {2\pi m n \over W} \right) \, ,~~~
\alpha=2,3 .
\eeq
Using this matrix, scalar variable $x_{\bn}$ defined on the site $\bn$ is 
transformed as
\beq
{x}_{n_1}^{({\bm m})} = {2 \over W}
\sum_{n_2, n_3} \cos \left( {2\pi m_2 n_2 \over W }\right)
\cos \left( { 2\pi m_3 n_3 \over W } \right) x_{\bn} ,~~
\eeq
where the vector ${\bm m}$ stands for ${\bm m}=(m_2, m_3)$, ($m_2,m_3=1,...,W$). 
By this transformation, 
the Hamiltonian with $\Delta=0$ is transformed into  $W^2$ independent chains
of the form ${\cal H}= \sum_{\bm m}\sum_{n_1} \left[
({\dot{x}}^{(\bm m )}_{n_1} )^2  +
{{x}}^{(\bm m )}_{n_1}  {\bm K}^{({\bm m})}_{n_1 , n_1 '} {{x}}^{(\bm m )}_{n_1'} \right] /2
$,
where ${\bm K}^{({\bm m})}_{n_1 , n_1 '}$ is  
the ${\bm m}$-mode representation of the force matrix ${\bm K}$:
\beq
{\bm K}^{({\bm m})}_{n_1 , n_1 '}
= \left\{
\begin{array}{l}
2  + k_0 + 4 \left[ \sin^2 \left( {\pi m_2 \over W}\right)
+ \sin^2 \left( {\pi m_3 \over W}\right)
 \right]
  \\
 ~~~~~~~~~~~~~~~~~~~~~~ \cdots ~ n_1=n_1' \\
-1 ~~~~~~~~~~~~~~~~~~ \cdots ~
|n_1 - n_1'|=1
\end{array} 
\right. .~~~
\eeq 
In addition, noise terms preserve the correlations 
\beq
\langle {\eta}_{\ell_1}^{({\bm m})} (t)  {\eta}_{\ell_1' }^{({\bm m} ')}  (t')
\rangle 
&=& 2 \gamma \, T_L \,\delta_{\ell_1 , \ell_1 '} \delta_{{\bm m}, {\bm m}'} \delta (t - t' ),~~ \\
\langle {\eta}_{r_1}^{({\bm m})} (t)  {\eta}_{r_1' }^{({\bm m} ')}  (t') 
\rangle 
&=& 2 \gamma \, T_R \,\delta_{r_1 , r_1 '} \delta_{{\bm m}, {\bm m}'} \delta (t - t' ) . ~~
\eeq
Hence, not only Hamiltonian but also Langevin equations are decomposed into $W^2$ independent
Langevin dynamics. 

We now consider the transmission matrix in terms of  which the CGF can be
written. 
The transmission matrix is diagonalized into the ${\bm m}$-mode 
\beq
{\cal T}^{({\bm m})} &=& 4 \gamma^2 \omega^2 |{\bm G}^{+\,({\bm m})}_{1, N} |^2 \, . 
\eeq
with the Green's function given by the inverse of the tridiagonal matrix 
\beq
\Bigl[\left[ {\bm G}^{+\,({\bm m})} \right]^{-1}\Bigr]_{n_1 , n_1 '}
&=& -\omega^2 - {\bm K}^{({\bm m})}_{n_1 , n_1 '} \nonumber \\
&-&i\gamma \omega \delta_{n_1, n_1 '}
(\delta_{n_1,1} + \delta_{n_1, N} ) . 
\eeq
The expression of ${\bm G}^{+\,({\bm m})}_{1, N}$ is readily obtained, 
hence, we finally get the explicit formula of $\mu_{ HC} (\lambda )$ for ordered harmonic crystal as follows.
\beq
\mu_{ HC} (\lambda )
&=& 
-{1\over 2\pi} \sum_{\bm m}
\int_{0}^{\infty} d\omega 
\log 
\Bigl[
 \nonumber \\
&& ~~1 -{4 \gamma^2 \omega^2 \sin^2 \theta_{\bm m} 
\over |\Lambda_{\bm m} |^2} 
T_L T_R \lambda (\lambda + \Delta\beta ) 
\Bigr] ,~~~~~~   
\eeq
where
\beq
\cos \theta_{\bm m} &=& \!
{ 2  + k_0 - \omega^2 \over 2}
+ 2 \left[ \sin^2 \left( {\pi m_2 \over W}\right)
+ \sin^2 \left( {\pi m_3 \over W}\right)
 \right]   , \nonumber \\
\Lambda_{\bm m} &=& 
\left[ \left( 1 - \gamma^2 \omega^2   \right) \cos\theta_{\bm m} - 2 i \gamma \omega \right] \sin (N \theta_{\bm m} )  \nonumber \\
&& + \left( 1 + \gamma^2 \omega^2 \right) \sin\theta_{\bm m}
\cos (N\theta_{\bm m} ) \,
. \nonumber 
\eeq

\end{document}